%% file: main.tex
\documentclass[conference]{IEEEtran}
\usepackage{tikz}
\usepackage{xcolor}
\usepackage{pgfplots}
\usepackage{pgfplotstable}
\usepgfplotslibrary{groupplots}
\usepgfplotslibrary{colormaps}
\pgfplotsset{compat=1.18}
\usetikzlibrary{patterns}

\usepackage{float}
\usepackage{algorithm}
\usepackage{algpseudocode}
\usepackage[ruled,vlined,linesnumbered,algo2e]{algorithm2e}

\DontPrintSemicolon
\SetAlgoNlRelativeSize{-1}
\SetAlFnt{\small}
\SetInd{0.6em}{1.0em}     
\SetKwInput{KwIn}{Require}
\SetKwInput{KwOut}{Ensure}
\SetKwProg{Fn}{Function}{:}{}
\SetKw{Return}{return}

\usepackage{eso-pic}

\newcommand{\placecopyrightnotice}{%
\AddToShipoutPictureFG*{%
  \AtPageLowerLeft{%
    \raisebox{0.45in}{%
      \hspace{0.55in}%
      \parbox[b]{\dimexpr\paperwidth-1.10in\relax}{%
        \centering
        \fontsize{6.5}{7.2}\selectfont
        \copyright 2026 IEEE. Personal use of this material is permitted. Permission from IEEE must be obtained for all other uses,\\
        in any current or future media, including reprinting/republishing this material for advertising or promotional purposes,\\
        creating new collective works, for resale or redistribution to servers or lists, or reuse of any copyrighted component of this work in other works.%
      }%
    }%
  }%
}%
}

\usepackage{enumitem}
\setlist[itemize]{leftmargin=*}

\input{pgfplots/pgfplots_setup}

\usepackage{mathcomSTEv4}
\usepackage{amsmath}
\usepackage{amsthm}
\usepackage{amssymb}
\usepackage{amsfonts}
\usepackage{bm}
\usepackage{cancel}
\usepackage[overload]{empheq} 
\usepackage{fix-cm} 

\usepackage{tabularx}
\usepackage{longtable} 
\usepackage{colortbl}

\usepackage{cite}

\IEEEoverridecommandlockouts

\title{\LARGE \bf
JGRA: Jacobian Geometry Robustness Assessment in\\
NISQ Noise-Aware Quantum Neural Networks
}

\author{Gianluca Scanu, Luca Barletta, and Stefano Rini
 \thanks{G. Scanu and L. Barletta are with the Dipartimento di Elettronica, Informazione e Bioingegneria, Politecnico di Milano, 20133 Milan, Italy.
     {\tt\small 
     gianluca.scanu@mail.polimi.it \newline
     luca.barletta@polimi.it}}%
\thanks{S. Rini is with the Department of Electrical and Computer Engineering, National Yang Ming Chiao Tung University (NYCU), Hsinchu, Taiwan. 
     {\tt\small stefano.rini@nycu.edu.tw}}%
}

\definecolor{myDarkGreen}{RGB}{0,100,0} 
\definecolor{myDarkBlue}{RGB}{0,0,139} 
\definecolor{myDarkGray}{gray}{0.25} 

\begin{document}

\maketitle
\placecopyrightnotice

\thispagestyle{empty}
\pagestyle{empty}

\begin{abstract}
The NISQ era places stringent constraints on quantum computation, where noise and decoherence fundamentally limit performance. In classical deep learning, model robustness and resilience to perturbations are well studied: deep neural networks (DNNs) maintain high performance despite pruning, noise injection, and structural perturbations due to inherent redundancy in their representations. A central challenge in quantum machine learning is to transfer this notion of robustness to quantum neural networks (QNNs) under realistic NISQ noise.
While classical deep learning exhibits robustness through structural redundancy, analogous principles for QNNs remain underdeveloped. 
We propose JGRA: a framework for assessing robustness in noise-aware QNNs via Jacobian geometry, capturing model sensitivity to parameter perturbations induced by noise. 
Our method includes entropy-matched noise calibration, noise-aware training, and noise-conditioned Jacobian extraction, yielding geometric descriptors that link clean-regime structure to noisy inference behaviour. 
We also empirically demonstrate that these descriptors encode predictive information about robustness under unseen noise.
\end{abstract}

\section{Introduction}

\noindent Quantum computing in the NISQ regime is fundamentally constrained by decoherence, gate imperfections, and limited circuit depth \cite{Preskill_2018}. 
In this setting, the central challenge is not only to design quantum algorithms, but to design algorithms that remain reliable under realistic noise. 
Explicitly robustifying every possible quantum procedure against hardware imperfections is impractical, as noise mechanisms are diverse and device-dependent. 

A promising alternative is offered by quantum machine learning (QML), where robustness can be partially ``baked into'' the training process. 
Variational quantum circuits and quantum neural networks (QNNs) are trained directly on noisy hardware or under calibrated noise models, enabling the model to adapt its internal representations to the physical constraints of the device \cite{Bharti_2022}. 
This paradigm shifts the burden of robustness from analytical circuit design to data-driven optimization, echoing the resilience observed in classical deep neural networks, which maintain functionality despite pruning, noise injection, and structural perturbations.
However, this shift raises a fundamental question: \emph{how should robustness of a trained QNN be assessed?} While performance degradation under noise can be empirically measured, it remains unclear whether structural properties of a trained model encode predictive information about its resilience to unseen perturbations. 
In this work, we take a first step toward answering this question by introducing a geometry-based framework for assessing robustness in noise-aware QNNs. 

\noindent
\textbf{State of the Art:}
Robustness and trainability of variational quantum models in the NISQ regime have been primarily studied through three interconnected lines of work.

First, trainability limitations have been formalized through the barren plateau phenomenon, where gradients vanish exponentially with system size or circuit depth, rendering optimization impractical for broad classes of parametrized circuits \cite{McClean_2018}. Subsequent analyses have shown that noise can exacerbate this effect, further suppressing gradient variance and accelerating the onset of flat landscapes \cite{Larocca_2025, Wang_2021}. 

Second, noise-aware and noise-mitigation strategies have been proposed to improve empirical robustness. These include explicit noise injection during training, Pauli-twirling approximations, and hybrid error mitigation approaches \cite{Wang_2021, wallmantwirling}. While such techniques demonstrate improved performance under specific noise models, robustness is typically evaluated through output-level metrics such as accuracy degradation, without analyzing how noise modifies the internal geometric structure of the model.

Third, a geometric perspective on quantum learning has emerged. The quantum geometric tensor and quantum Fisher information define natural metrics in parameter space \cite{Stokes_2020, Meyer_2021}, and the quantum neural tangent kernel links training dynamics to spectral properties of the induced geometry \cite{Liu_2022}. These works highlight that trainability and sensitivity are intrinsically geometric phenomena and motivate a geometric view of robustness: noise affects performance through deformations of the trained input-output map. However, the connection between such geometric descriptors and robustness under realistic noise exposure remains largely unexplored.

Collectively, existing studies show that noise alters optimization dynamics, that noise-aware training can improve empirical resilience, and that geometric properties govern learning behaviour. 

\noindent
\textbf{Contribution:}
We propose a geometry-based framework for assessing robustness of noise-aware quantum neural networks in the NISQ regime. The framework combines (i) entropy-matched noise calibration for cross-channel comparability, (ii) noise-aware training with deterministic, decoupled Jacobian extraction, (iii) a suite of static, stability, and physics-informed alignment descriptors of parameter-space sensitivity, and (iv) a robustness probing procedure that evaluates whether clean-regime geometric structure predicts performance under unseen inference noise.

\section{System Model: NA-QNN}
\label{sec:System Model}

We consider supervised learning models implemented as variational QNNs. A QNN defines a parameterized input–output map
$f_{\boldsymbol{\theta}} : \mathcal{X} \rightarrow \mathbb{R}^k$, where $\boldsymbol{\theta} \in \mathbb{R}^P$ denotes the vector of trainable quantum parameters.

\smallskip
\noindent
\textbf{Encoding.} Classical input data $x \in \mathcal{X}$ are embedded into a quantum state through a fixed encoding map
\begin{equation}
\mathcal{E} : x \mapsto \rho(x),
\qquad
\rho(x) \in \mathcal{D}(\mathcal{H}),
\end{equation}
where $\mathcal{D}(\mathcal{H})$ denotes the set of density operators on a $2^n$-dimensional Hilbert space $\mathcal{H}$.

\smallskip
\noindent
\textbf{Parameterized quantum evolution.}
The encoded state is processed by a parameterized quantum circuit
\begin{equation}
\rho_{\boldsymbol{\theta}}(x)
=
U_{\boldsymbol{\theta}} \rho(x) U_{\boldsymbol{\theta}}^\dagger,
\end{equation}
where $U_{\boldsymbol{\theta}}$ is composed of alternating fixed and trainable layers,
\[
U_{\boldsymbol{\theta}}
=
W_L V_L(\boldsymbol{\theta}_L) \cdots
W_1 V_1(\boldsymbol{\theta}_1),
\]
with $V_\ell(\boldsymbol{\theta}_\ell)$ denoting variational blocks and $W_\ell$ fixed circuit components.

\smallskip
\noindent
\textbf{Measurement and output.}
The classical output is obtained via expectation values of fixed Hermitian observables $\{O_j\}_{j=1}^k$,
\begin{equation}
f_{\boldsymbol{\theta}}(x)
=
\left(
\langle O_1 \rangle_{\boldsymbol{\theta},x},
\ldots,
\langle O_k \rangle_{\boldsymbol{\theta},x}
\right),
\quad 
\langle O_j \rangle_{\boldsymbol{\theta},x}
=
\mathrm{Tr}\!\left[
O_j \rho_{\boldsymbol{\theta}}(x)
\right].    
\end{equation}

\smallskip
\noindent
\textbf{Noise model.}
In the NISQ setting, quantum evolution is affected by noise. We model noisy execution through a completely positive trace-preserving (CPTP) map $\mathcal{N}_\epsilon$, parameterized by a noise strength $\epsilon$, acting either during training or inference. The noisy model is therefore described by
\[
\rho_{\boldsymbol{\theta}}^{(\epsilon)}(x)
=
\mathcal{N}_\epsilon
\big(
U_{\boldsymbol{\theta}} \rho(x) U_{\boldsymbol{\theta}}^\dagger
\big),
\]
leading to noise-dependent outputs $f_{\boldsymbol{\theta}}(x;\epsilon)$.

When noise is explicitly integrated into the training procedure, we refer to the resulting model as a \emph{noise-aware quantum neural network (NA-QNN)}. In this work, we analyze the sensitivity and robustness properties of the trained NA-QNN mapping $f_{\boldsymbol{\theta}^\star}(x;\epsilon)$ through the geometry induced by its Jacobian with respect to $\boldsymbol{\theta}$.

\section{Proposed Framework: JGRA   }
\label{sec:framework}

We introduce the \emph{Jacobian Geometry Robustness Assessment (JGRA)} framework, a modular procedure for analyzing robustness of NA-QNNs as introduced in Sec.~\ref{sec:System Model}.
The JGRA framework consists of three components: (i) entropy-matched noise calibration, (ii) noise-conditioned Jacobian extraction, and (iii) geometry-based robustness descriptors.

\smallskip
\noindent
\textbf{Entropy-matched noise calibration.}
Different quantum noise channels are parameterized in incompatible ways; identical parameter values across channels do not correspond to equivalent physical perturbations. To enable principled cross-channel comparison, we calibrate noise strength using the entropy of a quantum channel \cite{Gour_2021}. Other calibration scales, such as average infidelity, output variance matching, or diamond distance, could emphasize different aspects of noise strength and remain future work.
For a CPTP map $\mathcal{E}$, the channel entropy is defined as
\begin{align}
H(\mathcal{E}) &
=
\log_{2} |B| \nonumber\\
-
& \sup_{\psi_{RA}}
D\!\left(
(\mathrm{id}_{R} \otimes \mathcal{E})
\bigl(\lvert \psi_{RA} \rangle \langle \psi_{RA} \rvert\bigr)
\;\big\|\;
\rho_{R} \otimes \pi_{B}
\right)    
\end{align}
where the supremum is taken over all purifications  $\lvert \psi_{RA} \rangle$ of input states~\cite{UHLMANN1976273}, and
$D(\cdot\|\cdot)$ denotes the quantum relative entropy~\cite{Wilde_2013}. Here, $\rho_R = \operatorname{Tr}_A\!\left[\lvert \psi_{RA} \rangle \langle \psi_{RA} \rvert\right]$ indicates the reduced state, while $\pi_B = \mathbb{I}/|B|$ is the maximally mixed state on the output system~\cite{UHLMANN1976273}.
Noise parameters are calibrated such that different channel families share equal entropy levels,
\[
H(\mathcal{E}^{a}_{\epsilon_1})
=
H(\mathcal{E}^{b}_{\epsilon_2}),
\]
yielding a common scalar
notion of noise strength independent of channel structure. The formal algorithmic description appears in Appendix.

\smallskip
\noindent
\textbf{Noise-conditioned Jacobian extraction.}
Given a trained model with parameters $\boldsymbol{\theta}^\star$, robustness is analyzed through the Jacobian of the noisy input–output mapping
\[
g_i(\epsilon)
=
\frac{\partial}{\partial \boldsymbol{\theta}}
f(x_i;\boldsymbol{\theta}^\star,\epsilon)
\in \mathbb{R}^P, \quad i=1,\dots, N.
\]

Per-sample Jacobians are stacked into
\[
J(\epsilon) \in \mathbb{R}^{N \times P}.
\]
Extraction is deterministic and performed at shared inference operating points, enabling direct comparison across trained models.

\smallskip
\noindent
\textbf{Geometry descriptors.} We consider three quantities: 

\noindent
\emph{Static geometry.}
The clean-regime Jacobian ($\epsilon=0$) induces the Gram matrix
\begin{equation}
G = \frac{1}{N} J(0)^\top J(0),
\end{equation}
which defines a Riemannian metric on parameter space. The clean Jacobian isolates how noise-aware training reshapes the learned sensitivity map, and spectral properties of $G$ characterize the intrinsic sensitivity structure.

\noindent
\emph{Stability under perturbations.}
Since small perturbations satisfy $\Delta f \approx J \Delta \theta$ and the Gram matrix $G$ captures the average first-order perturbation, the local deformation of geometry under noise perturbations is quantified through
\begin{equation}
r_i(\epsilon)
=
\frac{\|g_i(\epsilon+\delta) - g_i(\epsilon)\|_2}
{\|g_i(\epsilon)\|_2 + \eta},
\end{equation}
and via principal-angle variations between dominant eigenspaces of $G(\epsilon)$.

\noindent
\emph{Noise alignment.}
To relate geometry to physical noise structure, Pauli-transfer representations of channels are lifted into a diagonal parameter-space proxy $\Sigma_{\text{noise}}$. 
Alignment quantifies whether learned sensitivity directions coincide with those affected by noise through
\begin{equation}
\mathrm{Align}
=
\frac{\mathrm{Tr}(G\,\Sigma_{\text{noise}})}
{\mathrm{Tr}(G)\,\mathrm{Tr}(\Sigma_{\text{noise}})}.
\end{equation}

\begin{figure*}[t]
    \centering

    \includegraphics[width=0.78\textwidth,height=0.28\textheight]{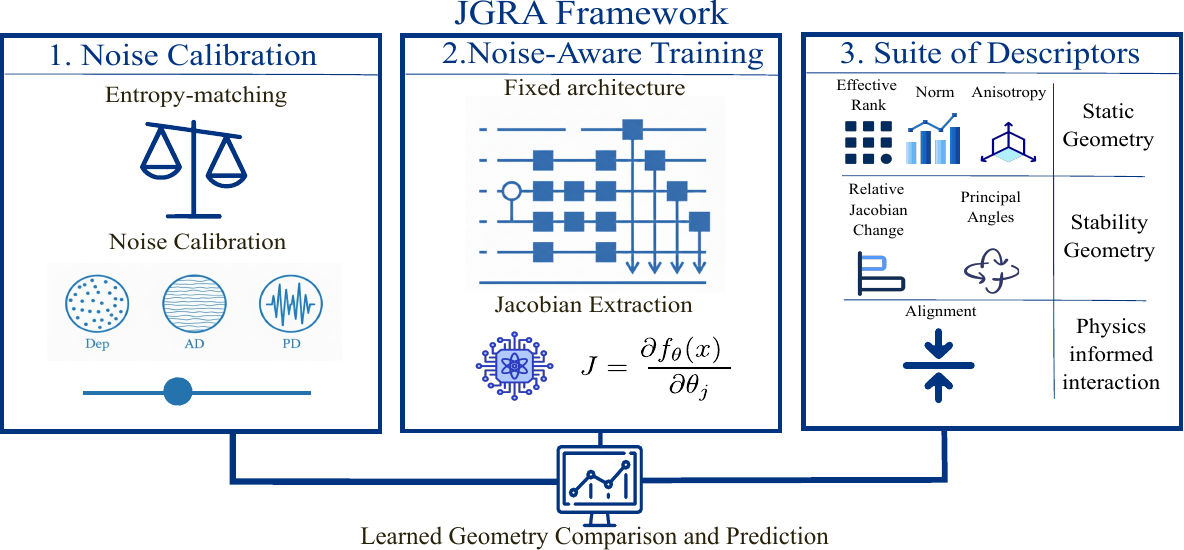}
    \caption{Overview of the noisy geometry evaluation framework. A model zoo is constructed by fixing the variational circuit and by training it under different entropy-matched noise-aware regimes. Jacobians with respect to quantum parameters are extracted in the clean inference regime and under small noise perturbations for each trained model, obtaining a Fisher-like geometry in parameter space. Static, stability, and physics-informed alignment descriptors are derived from this extraction process and later used for inference-time robustness prediction.}
    \label{fig:framework}
\end{figure*}

\medskip
For completeness, the full procedure is reported in Appendix. A graphical representation of the JGRA framework is presented in Fig.~\ref{fig:framework}.

\section{Experimental Setup}

\noindent\textbf{Model and data.}
We implement the variational model in PennyLane~\cite{bergholm2022pennylaneautomaticdifferentiationhybrid}. Inputs are amplitude-encoded on $n=4$ qubits. The ansatz is hardware-efficient with $L=3$ repeated layers of single-qubit rotations and entangling operations, for a total of $P=4\times2\times3=24$ trainable quantum parameters. Experiments use a PCA-preprocessed MNIST binary task (digits 0 vs.\ 1), consistently with training.

\smallskip

\noindent\textbf{Training noise and entropy matching.}
Training-time noise is injected via depolarizing (Dep), amplitude damping (AD), phase damping (PD), and their sequential compositions. Because channel parameters are not directly comparable, strengths are calibrated by entropy matching (Sec.~III): fixing a reference probability $p_{\mathrm{ref}}=0.1$, channel-specific probabilities are obtained by grid search over $p\in[10^{-4},0.5]$ (500 points). We consider three magnitudes $m\in\{0.5,1.5,3.0\}$, yielding 21 noise-aware training configurations.

\smallskip

\noindent\textbf{Jacobian extraction and stability probes.}
All Jacobians are computed \emph{post-training} with frozen parameters. We extract Jacobians on a deterministic validation subset of size $N_{\mathrm{val}}=400$ from the MNIST test split, balanced across classes via slicing. For each inference noise level $\epsilon$, we compute per-sample gradients of the class-1 probability,
$g_i(\epsilon)=\partial_{\boldsymbol{\theta}} f(x_i;\boldsymbol{\theta}^\star,\epsilon)$,
and stack them row-wise into $J(\epsilon)\in\mathbb{R}^{N\times P}$.
Stability is evaluated using a fixed \emph{Triple} inference channel (Dep$+$AD$+$PD) calibrated by the same entropy-matching procedure, at inference baselines $\epsilon\in\{0.0,0.5,1.5\}$ with finite perturbation $\delta=0.05$.

\smallskip
\noindent\textbf{Robustness targets.}
Robustness is evaluated under four test-time channels (Dep, AD, PD, Triple) using accuracy curves over a fixed grid of inference noise strengths and a shared test set. From these curves we compute three targets: area under the robustness curve (AURC), collapse point $\lambda_{50}$, and total variation (TV).

\noindent \textbf{Signal detection framework setup.} The prediction task is formulated as a supervised regression problem, in which robustness targets are estimated through a fixed vector of features derived from those extracted from trained models. As this study is performed after Jacobian extraction and metrics analysis, all prediction experiments are carried out without any change to the model parameters or the Jacobians.

\smallskip
\noindent\textbf{Signal-detection regression protocol.}
Robustness forecasting is posed as supervised regression from Jacobian-derived features (static, stability, and alignment descriptors). Each training configuration is identified by (training noise type, training noise strength) and repeated over three random seeds, producing 63 trained models; pairing each model with four test-noise channels yields 252 prediction rows.
We evaluate signal using leave-one-configuration-out cross-validation (LOCO-CV): all rows from the same training configuration (across seeds and test-noise channels) form one fold to prevent leakage.
We test Ridge, Kernel Ridge (RBF kernels over geometry and stability, combined via an interaction kernel), and Random Forest regressors. Ridge regularization is selected over $\alpha\in\{0,10^{-3},10^{-2},10^{-1},1,10,100\}$; Kernel Ridge uses $\alpha=0.1$, $\gamma_g,\gamma_s\in\{0.01,0.1,1.0\}$, and $\lambda\in\{0.0,0.5,1.0\}$; Random Forest uses a representative grid over $n_{\text{estimators}}\in\{200,500\}$, $\text{max\_depth}\in\{2,3,5\}$, $\text{min\_samples\_leaf}\in\{2,5\}$, and $\text{max\_features}\in\{\text{sqrt},0.7\}$.

\section{Results}
\begin{figure}[t]
    \centering

    \includegraphics[width=0.9\columnwidth]{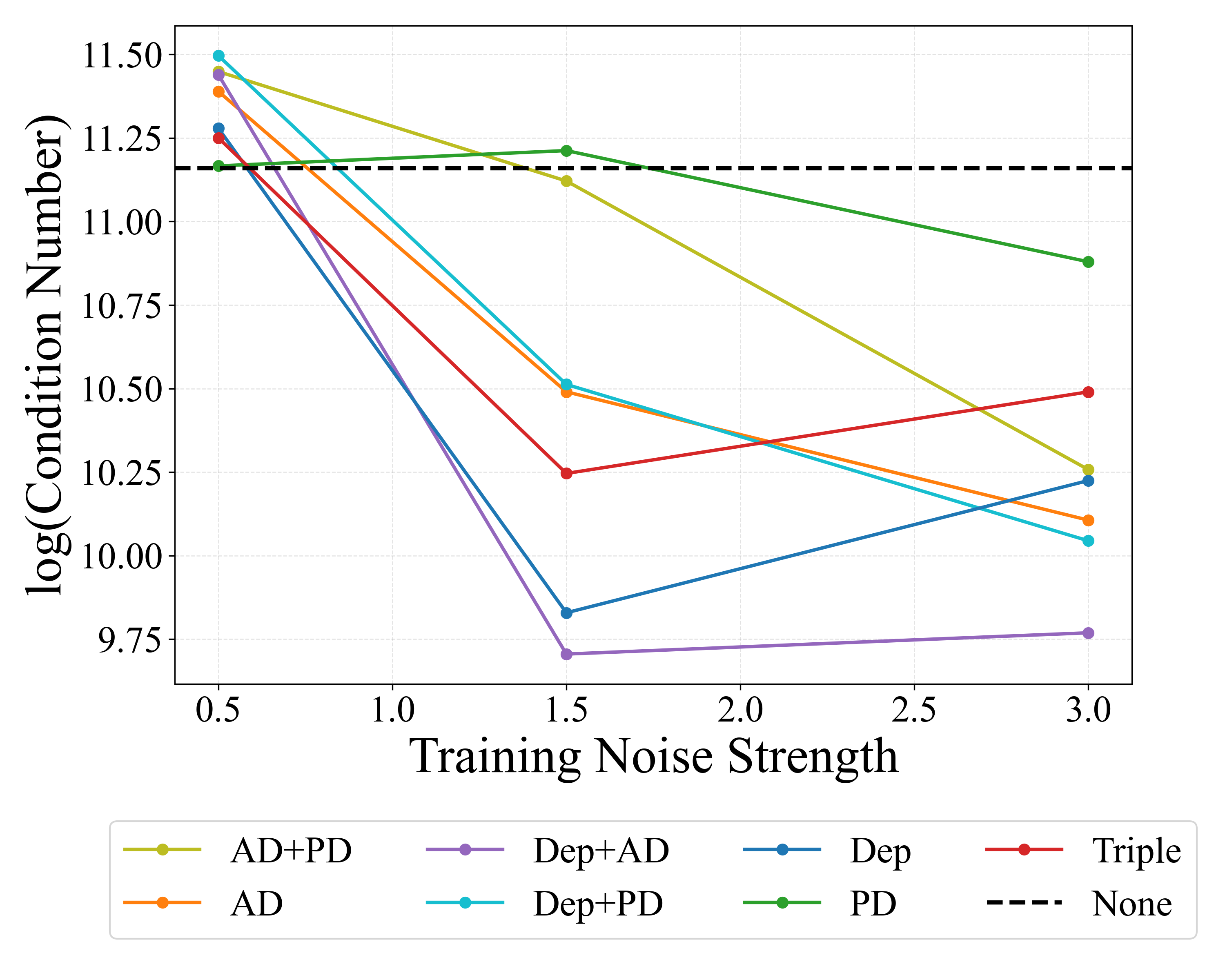}
    \caption{Logarithm of the condition number of $G$, defined as the ratio between the largest and smallest non-negligible eigenvalues, indicating the separation between extremal sensitivity directions in parameter space.}
    \label{fig:cond_vs_noise}
\end{figure}

\begin{figure}[t]
    \centering

    \resizebox{\columnwidth}{!}{%
    \input{pgfplots/fig_relative_deltaJ_stacked}%
    }
    \caption{Median relative Jacobian variation $\Delta J_{\mathrm{rel}}$ under small inference-noise perturbations $\delta=0.05$, evaluated at fixed inference noise levels $\varepsilon = \{0.0, 0.5, 1.5\}$. Each panel reports $\Delta J_{\mathrm{rel}}$ as a function of training noise strength for different inference noise levels. The noiseless model is reported as a baseline reference.}
    \label{fig:deltaJ_relative}
\end{figure}
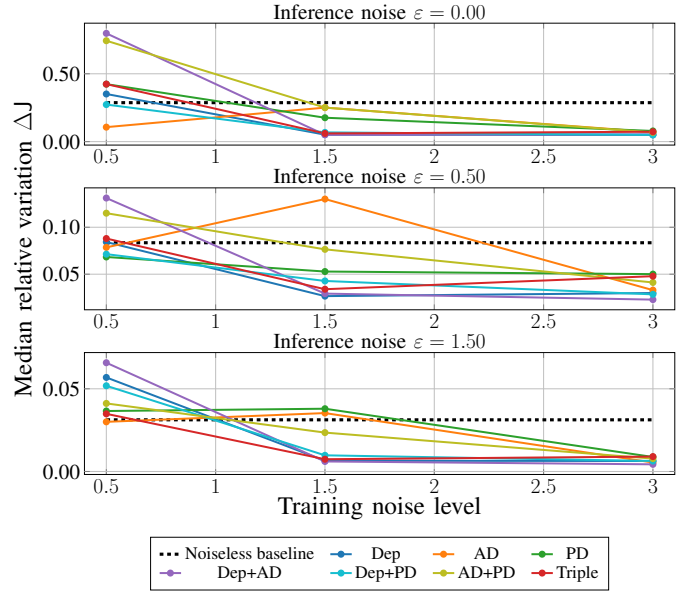

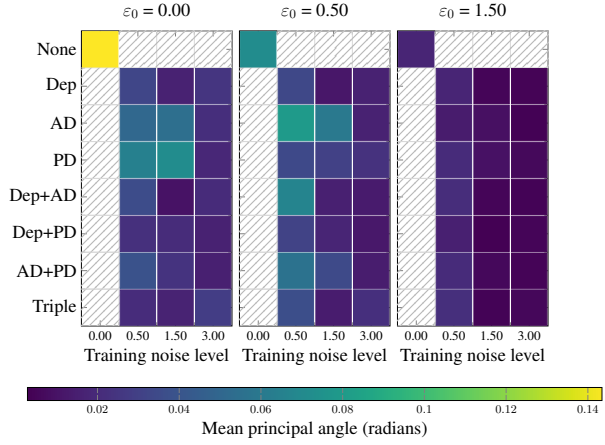
\begin{figure}[t]
    \centering

    \resizebox{0.9\columnwidth}{!}{%
    \input{pgfplots/fig_principal_angle_heatmap}%
  }
    \caption{Mean principal angles between dominant Jacobian eigenspaces under local inference-noise perturbations. Principal angles are computed between the leading eigenspaces of the model's Gram matrix $G(\varepsilon_0)$ and its perturbed version $G(\varepsilon_0 + \delta)$. In each panel, corresponding to a different inference operating point, darker colors indicate smaller angles and thus greater directional stability. Hatched entries denote configurations that are not defined or not meaningful for comparison, such as the noiseless model under increasing training noise strength.}
    \label{fig:principal_angles}
\end{figure}

\begin{figure}[t]
    \centering

    \resizebox{\columnwidth}{!}{%
    \input{pgfplots/fig_alignment}
    }
    \caption{PTM-based alignment index between the clean Jacobian geometry $G$ of each trained model and the parameter-space noise covariance $\Sigma_{noise}$ associated with the corresponding noise channel. Each of the reported curves corresponds to a specific noise family (AD, PD, Dep, and mixtures). Dashed horizontal lines indicate the alignment of the noiselessly trained model with each noise channel at the same inference level. Lower values indicate lower overlap, while larger ones indicate greater alignment between noise and model geometry.}
    \label{fig:align}
\end{figure}
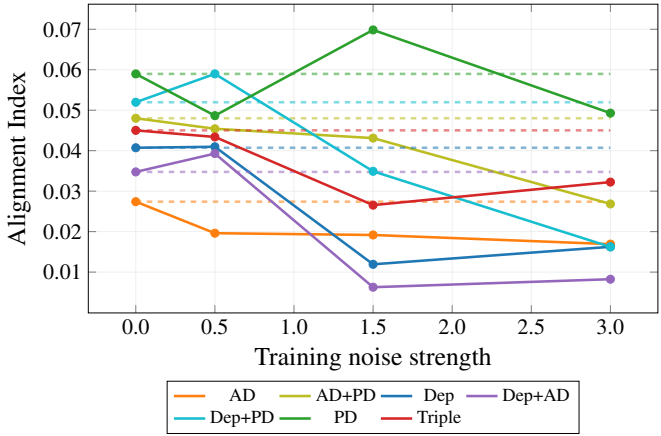

\begin{figure}[t]
    \centering

    \includegraphics[width=0.9\columnwidth]{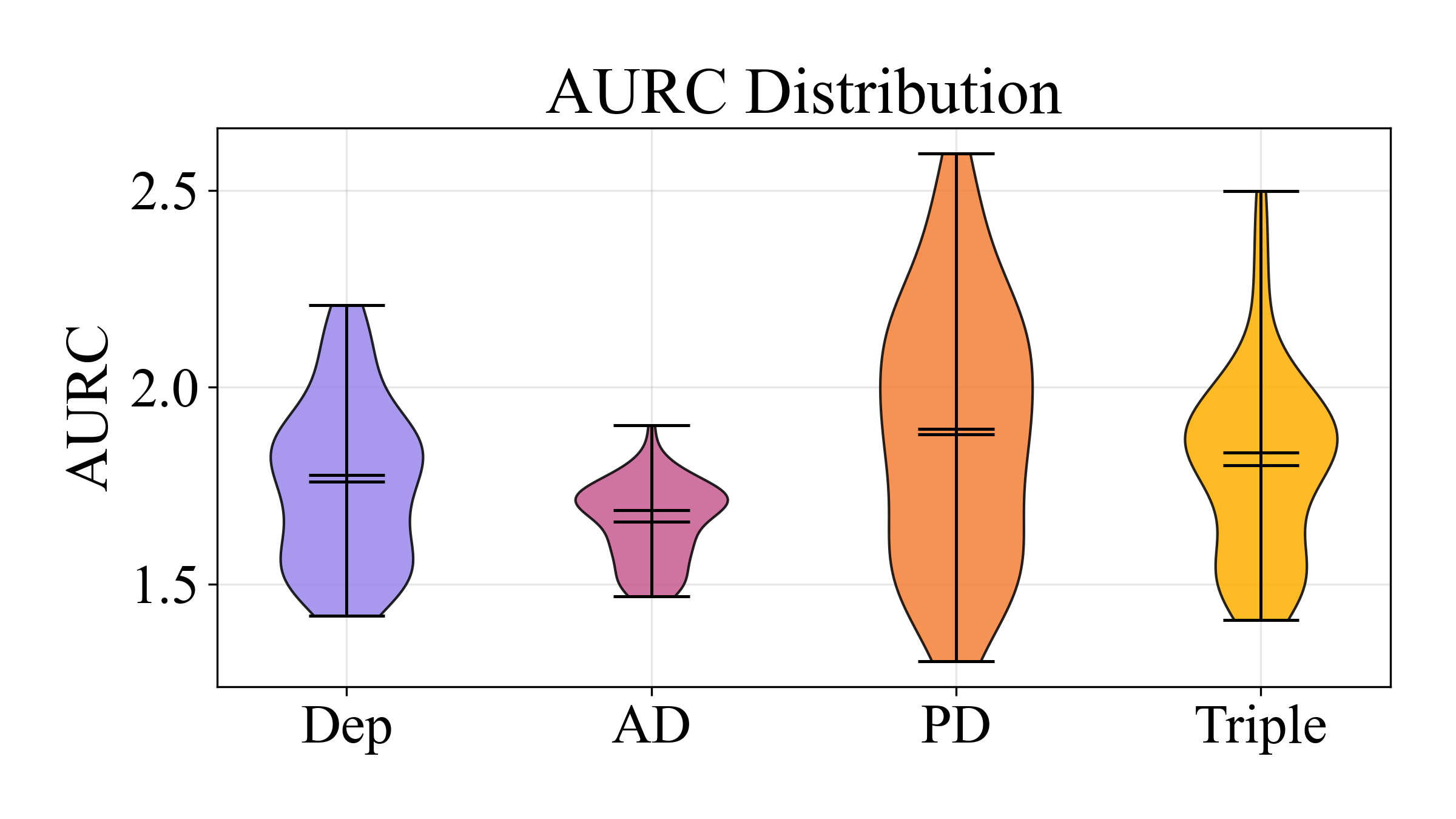}
    \caption{Empirical distributions through violin plots for the AURC target across test-time noise types. Each violin aggregates results across all training configurations, with horizontal markers indicating median and mean values.}
    \label{fig:violin}
\end{figure}

\begin{figure}[t]
    \centering

    \includegraphics[width=0.9\columnwidth]{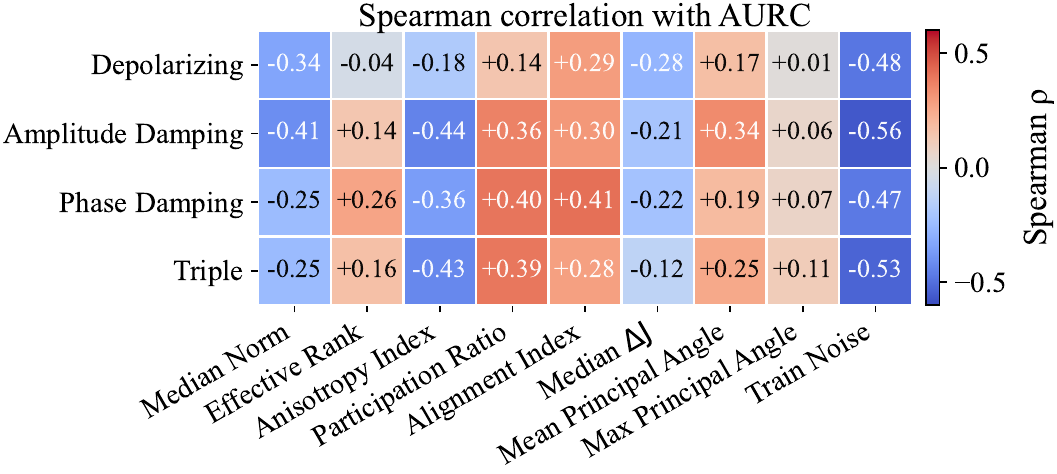}
    \caption{Spearman correlation coefficients between geometric features and AURC across test noise types (Depolarizing, Amplitude Damping, Phase Damping, and Triple).}
    \label{fig:corr_matrix}
\end{figure}

 \noindent\textbf{Static geometry.}
Across low-to-intermediate training-noise regimes, Jacobian norm distributions are right-skewed with heavy tails, indicating a small subset of highly sensitive samples. As training noise increases, tails are progressively truncated and mass shifts toward intermediate norms, suggesting that noise-aware training primarily suppresses extreme local sensitivities. Spectrally, the baseline Gram matrix shows a gradual eigenvalue decay (sensitivity spread across many directions), whereas high training-noise strengths compress the spectrum and strongly suppress eigenvalues. Intermediate regimes deviate from the baseline while preserving the hierarchy of leading modes, consistent with selective truncation of subdominant directions. Aggregate descriptors reflect these trends: the condition number decreases with training noise across channels (as pictured in Fig.~\ref{fig:cond_vs_noise}), while effective rank and related measures (e.g., fractional anisotropy vs.\ participation) vary in a channel-dependent way, indicating that improved conditioning does not correspond to a single universal mechanism.

\smallskip
\noindent\textbf{Stability under inference perturbations.}
We quantify local geometric deformation using (i) relative Jacobian variation under $\epsilon\mapsto\epsilon+\delta$ and (ii) principal-angle changes between dominant eigenspaces of $G(\epsilon_0)$ and $G(\epsilon_0+\delta)$. Relative variations decrease for noise-trained models across families and inference operating points (Fig.~\ref{fig:deltaJ_relative}), indicating reduced sensitivity to inference perturbations. Principal angles similarly shrink compared to the noiseless baseline (Fig.~\ref{fig:principal_angles}), with intermediate training-noise levels often yielding the most stable subspaces across $\epsilon_0\in\{0.0,0.5,1.5\}$.

\smallskip
\noindent\textbf{Noise alignment.}
To relate learned sensitivity directions to physical noise action, we compute the PTM-based alignment index between clean geometry $G$ and the channel proxy $\Sigma_{\text{noise}}$. Training under noise generally reduces geometry--noise overlap, consistent with a rotation/redistribution of dominant sensitivity away from noise-dominated directions (Fig.~\ref{fig:align}).

\smallskip
\noindent\textbf{Do clean descriptors contain robustness information?}
All descriptors are extracted at (or near) the clean operating point, while robustness targets (AURC, $\lambda_{50}$, TV) are computed under noisy inference; this separation tests whether robustness is partially encoded in intrinsic Jacobian structure rather than reflecting trivial corruption at extraction time. The AURC distribution is strongly noise-type dependent (PD and Triple are typically worse than Dep and AD) but exhibits substantial overlap across models (Fig.~\ref{fig:violin}), motivating model-level predictors. Pairwise Pearson/Spearman correlations (Fig.~\ref{fig:corr_matrix}) are generally weak-to-moderate and noise-dependent (often $|r|<0.5$), indicating the absence of a simple linear/monotone mapping; nevertheless, stable directional patterns emerge (e.g., AURC tends to decrease with median Jacobian norm and increase with alignment and degradation-related metrics). Similar heterogeneity holds for $\lambda_{50}$ and TV, with the largest spread often observed under PD and Triple.

\smallskip
\noindent\textbf{Signal-detection regression.}
Using LOCO-CV, a minimal Ridge baseline with training-noise strength alone explains limited but non-trivial variance in AURC (e.g., strongest under AD, $R^2 \approx 0.38$, weaker under Dep, $R^2 \approx 0.04$, and PD, $R^2 \approx 0.10$, with the Triple case at $R^2 \approx 0.26$), performs poorly on $\lambda_{50}$, and is competitive for TV mainly under PD ($R^2 \approx 0.47$). Ridge models using static geometry (G1) achieve comparable performance on AURC (notably under AD, $R^2 \approx 0.34$), and adding training-noise strength (G2) yields selective gains ($R^2$ rises to $\approx 0.42$ under AD for AURC and to $\approx 0.51$ under PD for TV), indicating complementarity between noise level and clean geometry. Linear models remain largely ineffective for $\lambda_{50}$, with near-zero or negative explained variance in most settings. Kernel Ridge, which probes nonlinear structure, yields regime-dependent gains and is most informative for $\lambda_{50}$, where it achieves positive explained variance in multiple channels (e.g., $R^2 \approx 0.14$ for PD and $R^2 \approx 0.28$ for Triple), suggesting that collapse thresholds depend on higher-order interactions among descriptors. Random Forest performance is generally modest and unstable in this dataset regime, reinforcing that the main outcome is evidence of a structured but weak, noise-dependent geometry — robustness signal — rather than high-accuracy forecasting.

\section{Discussion}
This work examined whether structural properties of NA-QNNs encode predictive information about robustness under noisy inference. 
The central question was whether clean-regime Jacobian geometry reflects intrinsic resilience, beyond mere performance measurements.
Static geometry descriptors reveal that noise-aware training reshapes parameter-space sensitivity in a structured manner. 
Increasing training noise regularizes Jacobian norms and compresses spectral spread, reducing extreme anisotropy and concentrating sensitivity along more stable dominant directions. This trend is not strictly monotonic: weak training noise can act as a perturbation before it acts as a regularizer, making noise strength a tunable hyperparameter.
Stability analyses further show that noise-trained models exhibit reduced geometric deformation under local inference perturbations, indicating smoother degradation under noise. Importantly, alignment-based metrics demonstrate that this stabilization is not isotropic: robustness depends on the specific noise channel, and no universal geometric signature of robustness emerges.
The predictive analysis confirms that geometry–robustness relationships are structured but modest in magnitude. Static geometry provides consistent improvements over noise-strength baselines, indicating that complementary information is encoded in clean Jacobians. 
However, linear models are insufficient to fully capture robustness-sensitive targets such as the collapse threshold $\lambda_{50}$, suggesting intrinsic nonlinear dependencies. 
Noise-specific patterns further highlight that AD exhibits the clearest geometry–robustness correspondence, while Dep noise—being isotropic—naturally limits discriminative geometric signal. Overall, robustness appears fundamentally noise-dependent rather than governed by a single universal descriptor.

These findings should be interpreted within scope limitations. The study considers a finite set of architectures and calibrated noise models, and robustness is measured through global degradation metrics rather than hardware-specific or long-time dynamical effects. Moreover, predictive analysis is exploratory and constrained by dataset size.

\section{Conclusion}

We introduced the Jacobian Geometry Robustness Assessment (JGRA) framework for assessing robustness of noise-aware quantum neural networks. 
By combining entropy-matched noise calibration, deterministic noise-conditioned Jacobian extraction, and complementary geometric descriptors, we provided a systematic methodology to relate parameter-space sensitivity structure to noisy inference behaviour.

Our results indicate that clean-regime Jacobian geometry contains structured, noise-dependent information about robustness, but that no universal robustness signature exists across noise channels. This work establishes a principled foundation for geometry-based robustness assessment in the NISQ regime.

Future research directions include extending the framework to broader architectural families, analyzing the temporal evolution of Jacobian geometry during training, incorporating richer noise models (e.g., correlated or non-Markovian effects), and learning geometry descriptors directly through representation-learning techniques.

\appendix

\section{Algorithms}
This appendix reports the pseudocode for (i) the entropy-matched
noise calibration used to equate different noise families to a common target entropy, and (ii) the PTM-based geometry--noise alignment procedure used to
derive $\Sigma_{\text{noise}}$ and the alignment index Align.

\begin{algorithm}[H]
\caption{Entropy Matching Algorithm (from Gour-Wilde operational channel entropy~\cite{Gour_2021})}
\label{alg:entropy_matching}
\textbf{Require:} noise family $\mathcal F$; target entropy $H^\star$; grid $\mathcal P\subset[p_{\min},p_{\max}]$; restarts $n_{\text{tries}}$\;\\
\textbf{Ensure:} matched parameter $p^\star$ s.t. $H(\mathcal E_{p^\star}^{\mathcal F})\approx H^\star$\;

\textbf{Subroutine} \textsc{ChannelEntropy}$(\mathcal E)$:\;
\Indp
$D_{\max}\leftarrow-\infty$;\quad $\pi_B\leftarrow I/2$;\quad $\log_2|B|\leftarrow 1$\;\\
Initialize $\mathcal S\leftarrow\{|\Phi\rangle\}\cup\{\text{random }|\psi_{RA}\rangle\}^{n_{\text{tries}}-1}$\;\\
\ForEach{$|\psi_{RA}\rangle\in\mathcal S$}{
  maximize $D(\sigma_{RB}\,\|\,\rho_R\otimes\pi_B)$ over $|\psi_{RA}\rangle$ \\
  $\sigma_{RB}\leftarrow(\mathrm{id}_R\otimes\mathcal E)(|\psi_{RA}\rangle\langle\psi_{RA}|)$\quad\\
  $\rho_R\leftarrow\mathrm{Tr}_A(|\psi_{RA}\rangle\langle\psi_{RA}|)$\;\\
  $D_{\max}\leftarrow\max(D_{\max},D(\sigma_{RB}\,\|\,\rho_R\otimes\pi_B))$\;
}
\textbf{return} $H(\mathcal E)\leftarrow \log_2|B|-D_{\max}$\;
\Indm

\ForEach{$p\in\mathcal P$}{
  $H(p)\leftarrow \textsc{ChannelEntropy}(\mathcal E_p^{\mathcal F})$\;
  $J(p)\leftarrow |H(p)-H^\star|$\;
}
$p^\star\leftarrow \arg\min_{p\in\mathcal P} J(p)$\;
\textbf{return} $p^\star$ (and $H(p^\star)$)\;
\end{algorithm}

\begin{algorithm}[H]
\caption{PTM-based Geometry-Noise Alignment \\ Procedure}
\label{alg:ptm_alignment}
\textbf{Require:} single-qubit channel $\mathcal E$; effective scale $p$; clean Jacobian $J\in\mathbb{R}^{N\times P}$; $\{P_j\}_{j=1}^P$, $P_j\in\{X,Y,Z\}$\;\\
\textbf{Ensure:} $\Sigma_{\mathrm{noise}}$ and $\mathrm{Align}$\;

$T(\mathcal E_p)\leftarrow$ PTM of $\mathcal E_p$ in basis $\{I,X,Y,Z\}$\;
$\Delta\leftarrow T(\mathcal E_p)-I_4$\;
\ForEach{$A\in\{X,Y,Z\}$}{
  $\beta_A\leftarrow \|\Pi\,\Delta e_A\|_2^2$\;
}
$\Sigma_{\mathrm{noise}}\leftarrow \mathrm{diag}(\beta_{P_1},\ldots,\beta_{P_P})$\;\\
$G\leftarrow \frac{1}{N}J^\top J$\;\\
$\mathrm{Align}\leftarrow \dfrac{\mathrm{Tr}(G\,\Sigma_{\mathrm{noise}})}{\mathrm{Tr}(G)\,\mathrm{Tr}(\Sigma_{\mathrm{noise}})}$\;\\
\textbf{return} $\Sigma_{\mathrm{noise}},\mathrm{Align}$\;
\end{algorithm}

\bibliographystyle{IEEEtran}
\bibliography{bibliography.bib}

\end{document}

%% file: pgfplots/pgfplots_setup.tex
\pgfplotstableread[col sep=comma]{pgfplots/csv_heat_delta/week5_jacobians_PennyLaneMNIST_noiseNone_trainNoise0.00.csv}\TNone

\pgfplotstableread[col sep=comma]{pgfplots/csv_heat_delta/week5_jacobians_PennyLaneMNIST_noiseDep_trainNoise0.50.csv}\TDepA
\pgfplotstableread[col sep=comma]{pgfplots/csv_heat_delta/week5_jacobians_PennyLaneMNIST_noiseDep_trainNoise1.50.csv}\TDepB
\pgfplotstableread[col sep=comma]{pgfplots/csv_heat_delta/week5_jacobians_PennyLaneMNIST_noiseDep_trainNoise3.00.csv}\TDepC

\pgfplotstableread[col sep=comma]{pgfplots/csv_heat_delta/week5_jacobians_PennyLaneMNIST_noiseAD_trainNoise0.50.csv}\TADA
\pgfplotstableread[col sep=comma]{pgfplots/csv_heat_delta/week5_jacobians_PennyLaneMNIST_noiseAD_trainNoise1.50.csv}\TADB
\pgfplotstableread[col sep=comma]{pgfplots/csv_heat_delta/week5_jacobians_PennyLaneMNIST_noiseAD_trainNoise3.00.csv}\TADC

\pgfplotstableread[col sep=comma]{pgfplots/csv_heat_delta/week5_jacobians_PennyLaneMNIST_noisePD_trainNoise0.50.csv}\TPDA
\pgfplotstableread[col sep=comma]{pgfplots/csv_heat_delta/week5_jacobians_PennyLaneMNIST_noisePD_trainNoise1.50.csv}\TPDB
\pgfplotstableread[col sep=comma]{pgfplots/csv_heat_delta/week5_jacobians_PennyLaneMNIST_noisePD_trainNoise3.00.csv}\TPDC

\pgfplotstableread[col sep=comma]{pgfplots/csv_heat_delta/week5_jacobians_PennyLaneMNIST_noiseDepAD_trainNoise0.50.csv}\TDepADA
\pgfplotstableread[col sep=comma]{pgfplots/csv_heat_delta/week5_jacobians_PennyLaneMNIST_noiseDepAD_trainNoise1.50.csv}\TDepADB
\pgfplotstableread[col sep=comma]{pgfplots/csv_heat_delta/week5_jacobians_PennyLaneMNIST_noiseDepAD_trainNoise3.00.csv}\TDepADC

\pgfplotstableread[col sep=comma]{pgfplots/csv_heat_delta/week5_jacobians_PennyLaneMNIST_noiseDepPD_trainNoise0.50.csv}\TDepPDA
\pgfplotstableread[col sep=comma]{pgfplots/csv_heat_delta/week5_jacobians_PennyLaneMNIST_noiseDepPD_trainNoise1.50.csv}\TDepPDB
\pgfplotstableread[col sep=comma]{pgfplots/csv_heat_delta/week5_jacobians_PennyLaneMNIST_noiseDepPD_trainNoise3.00.csv}\TDepPDC

\pgfplotstableread[col sep=comma]{pgfplots/csv_heat_delta/week5_jacobians_PennyLaneMNIST_noiseADPD_trainNoise0.50.csv}\TADPDA
\pgfplotstableread[col sep=comma]{pgfplots/csv_heat_delta/week5_jacobians_PennyLaneMNIST_noiseADPD_trainNoise1.50.csv}\TADPDB
\pgfplotstableread[col sep=comma]{pgfplots/csv_heat_delta/week5_jacobians_PennyLaneMNIST_noiseADPD_trainNoise3.00.csv}\TADPDC

\pgfplotstableread[col sep=comma]{pgfplots/csv_heat_delta/week5_jacobians_PennyLaneMNIST_noiseTriple_trainNoise0.50.csv}\TTripleA
\pgfplotstableread[col sep=comma]{pgfplots/csv_heat_delta/week5_jacobians_PennyLaneMNIST_noiseTriple_trainNoise1.50.csv}\TTripleB
\pgfplotstableread[col sep=comma]{pgfplots/csv_heat_delta/week5_jacobians_PennyLaneMNIST_noiseTriple_trainNoise3.00.csv}\TTripleC

\definecolor{cDep}{HTML}{1F77B4}
\definecolor{cAD}{HTML}{FF7F0E}
\definecolor{cPD}{HTML}{2CA02C}
\definecolor{cDepAD}{HTML}{9467BD}
\definecolor{cDepPD}{HTML}{17BECF}
\definecolor{cADPD}{HTML}{BCBD22}
\definecolor{cTriple}{HTML}{D62728}

\newcommand{\GetCell}[4]{%
  \pgfplotstablegetelem{#2}{#3}\of#1%
  \edef#4{\pgfplotsretval}%
}
\newcommand{\HeatPoint}[4]{%
  \GetCell{#1}{#2}{principal_angle_mean}{\m}%
  \addplot[
    scatter, only marks,
    mark=square*,
    mark size=13.5pt,
    point meta=explicit,
    scatter/use mapped color={draw=mapped color, fill=mapped color}
  ] coordinates { (#3,#4) [\m] };
}

\newcommand{\HeatPanelRow}[1]{%
  \HeatPoint{\TNone}{#1}{0}{0}%

  \HeatPoint{\TDepA}{#1}{1}{1}\HeatPoint{\TDepB}{#1}{2}{1}\HeatPoint{\TDepC}{#1}{3}{1}%

  \HeatPoint{\TADA}{#1}{1}{2}\HeatPoint{\TADB}{#1}{2}{2}\HeatPoint{\TADC}{#1}{3}{2}%

  \HeatPoint{\TPDA}{#1}{1}{3}\HeatPoint{\TPDB}{#1}{2}{3}\HeatPoint{\TPDC}{#1}{3}{3}%

  \HeatPoint{\TDepADA}{#1}{1}{4}\HeatPoint{\TDepADB}{#1}{2}{4}\HeatPoint{\TDepADC}{#1}{3}{4}%

  \HeatPoint{\TDepPDA}{#1}{1}{5}\HeatPoint{\TDepPDB}{#1}{2}{5}\HeatPoint{\TDepPDC}{#1}{3}{5}%

  \HeatPoint{\TADPDA}{#1}{1}{6}\HeatPoint{\TADPDB}{#1}{2}{6}\HeatPoint{\TADPDC}{#1}{3}{6}%

  \HeatPoint{\TTripleA}{#1}{1}{7}\HeatPoint{\TTripleB}{#1}{2}{7}\HeatPoint{\TTripleC}{#1}{3}{7}%
}

\newcommand{\HatchedCell}[2]{%
  \addplot[
    fill=gray!6,
    draw=gray!30,
    line width=0.2pt,
    pattern=north east lines,
    pattern color=gray!55
  ] coordinates {
    (#1-0.5,#2-0.5)
    (#1+0.5,#2-0.5)
    (#1+0.5,#2+0.5)
    (#1-0.5,#2+0.5)
  };
}

\newcommand{\DrawInvalidCells}{%
  \foreach \xx in {1,2,3}{%
    \HatchedCell{\xx}{0}%
  }%

  \foreach \yy in {1,2,3,4,5,6,7}{%
    \HatchedCell{0}{\yy}%
  }%
}

%% file: pgfplots/fig_relative_deltaJ_stacked.tex

\begin{tikzpicture}

\begin{groupplot}[
  group style={
    group size=1 by 3,
    vertical sep=1.2cm,
    group name=dj
  },
  width=\textwidth,
  height=5cm,
  xmin=0.4, xmax=3.1,
  xtick={0.5,1.0,1.5,2.0,2.5,3.0},
  grid=major,
  title style={yshift=-6pt, font=\LARGE},
  tick label style={font=\LARGE},
  label style={font=\LARGE},
  scaled y ticks=false,
  yticklabel style={/pgf/number format/fixed,/pgf/number format/precision=2}
]

\nextgroupplot[
  title={Inference noise $\varepsilon=0.00$},
  ytick=\empty,
  yticklabels=\empty,
  extra y ticks={0.0,0.5},
  extra y tick labels={0.00,0.50},
  extra tick style={grid=major},
  legend to name=RelLegendDeltaJ,
  legend columns=4,
  legend style={
  draw,
  fill=white,
  font=\Large,
  /tikz/every even column/.append style={column sep=10pt},
  inner sep=6pt
  }
]

\GetCell{\TNone}{0}{deltaJ_rel_median}{\b}
\addplot[black, dashed, line width=2.5pt, mark=none]
coordinates {(0.5,\b) (1.5,\b) (3.0,\b)};
\addlegendentry{Noiseless baseline}

\GetCell{\TDepA}{0}{deltaJ_rel_median}{\yA}
\GetCell{\TDepB}{0}{deltaJ_rel_median}{\yB}
\GetCell{\TDepC}{0}{deltaJ_rel_median}{\yC}
\addplot[color=cDep, line width=1.6pt, mark=*]
coordinates {(0.5,\yA) (1.5,\yB) (3.0,\yC)};
\addlegendentry{Dep}

\GetCell{\TADA}{0}{deltaJ_rel_median}{\yA}
\GetCell{\TADB}{0}{deltaJ_rel_median}{\yB}
\GetCell{\TADC}{0}{deltaJ_rel_median}{\yC}
\addplot[color=cAD, line width=1.6pt, mark=*]
coordinates {(0.5,\yA) (1.5,\yB) (3.0,\yC)};
\addlegendentry{AD}

\GetCell{\TPDA}{0}{deltaJ_rel_median}{\yA}
\GetCell{\TPDB}{0}{deltaJ_rel_median}{\yB}
\GetCell{\TPDC}{0}{deltaJ_rel_median}{\yC}
\addplot[color=cPD, line width=1.6pt, mark=*]
coordinates {(0.5,\yA) (1.5,\yB) (3.0,\yC)};
\addlegendentry{PD}

\GetCell{\TDepADA}{0}{deltaJ_rel_median}{\yA}
\GetCell{\TDepADB}{0}{deltaJ_rel_median}{\yB}
\GetCell{\TDepADC}{0}{deltaJ_rel_median}{\yC}
\addplot[color=cDepAD, line width=1.6pt, mark=*]
coordinates {(0.5,\yA) (1.5,\yB) (3.0,\yC)};
\addlegendentry{Dep+AD}

\GetCell{\TDepPDA}{0}{deltaJ_rel_median}{\yA}
\GetCell{\TDepPDB}{0}{deltaJ_rel_median}{\yB}
\GetCell{\TDepPDC}{0}{deltaJ_rel_median}{\yC}
\addplot[color=cDepPD, line width=1.6pt, mark=*]
coordinates {(0.5,\yA) (1.5,\yB) (3.0,\yC)};
\addlegendentry{Dep+PD}

\GetCell{\TADPDA}{0}{deltaJ_rel_median}{\yA}
\GetCell{\TADPDB}{0}{deltaJ_rel_median}{\yB}
\GetCell{\TADPDC}{0}{deltaJ_rel_median}{\yC}
\addplot[color=cADPD, line width=1.6pt, mark=*]
coordinates {(0.5,\yA) (1.5,\yB) (3.0,\yC)};
\addlegendentry{AD+PD}

\GetCell{\TTripleA}{0}{deltaJ_rel_median}{\yA}
\GetCell{\TTripleB}{0}{deltaJ_rel_median}{\yB}
\GetCell{\TTripleC}{0}{deltaJ_rel_median}{\yC}
\addplot[color=cTriple, line width=1.6pt, mark=*]
coordinates {(0.5,\yA) (1.5,\yB) (3.0,\yC)};
\addlegendentry{Triple}

\nextgroupplot[
  title={Inference noise $\varepsilon=0.50$},
  ytick=\empty,
  yticklabels=\empty,
  extra y ticks={0.05,0.10},
  extra y tick labels={0.05,0.10},
  extra tick style={grid=major}
]

\GetCell{\TNone}{1}{deltaJ_rel_median}{\b}
\addplot[black, dashed, line width=2.5pt, mark=none]
coordinates {(0.5,\b) (1.5,\b) (3.0,\b)};

\GetCell{\TDepA}{1}{deltaJ_rel_median}{\yA}\GetCell{\TDepB}{1}{deltaJ_rel_median}{\yB}\GetCell{\TDepC}{1}{deltaJ_rel_median}{\yC}
\addplot[color=cDep, line width=1.6pt, mark=*]
coordinates {(0.5,\yA) (1.5,\yB) (3.0,\yC)};

\GetCell{\TADA}{1}{deltaJ_rel_median}{\yA}\GetCell{\TADB}{1}{deltaJ_rel_median}{\yB}\GetCell{\TADC}{1}{deltaJ_rel_median}{\yC}
\addplot[color=cAD, line width=1.6pt, mark=*]
coordinates {(0.5,\yA) (1.5,\yB) (3.0,\yC)};

\GetCell{\TPDA}{1}{deltaJ_rel_median}{\yA}\GetCell{\TPDB}{1}{deltaJ_rel_median}{\yB}\GetCell{\TPDC}{1}{deltaJ_rel_median}{\yC}
\addplot[color=cPD, line width=1.6pt, mark=*]
coordinates {(0.5,\yA) (1.5,\yB) (3.0,\yC)};

\GetCell{\TDepADA}{1}{deltaJ_rel_median}{\yA}\GetCell{\TDepADB}{1}{deltaJ_rel_median}{\yB}\GetCell{\TDepADC}{1}{deltaJ_rel_median}{\yC}
\addplot[color=cDepAD, line width=1.6pt, mark=*]
coordinates {(0.5,\yA) (1.5,\yB) (3.0,\yC)};

\GetCell{\TDepPDA}{1}{deltaJ_rel_median}{\yA}\GetCell{\TDepPDB}{1}{deltaJ_rel_median}{\yB}\GetCell{\TDepPDC}{1}{deltaJ_rel_median}{\yC}
\addplot[color=cDepPD, line width=1.6pt, mark=*]
coordinates {(0.5,\yA) (1.5,\yB) (3.0,\yC)};

\GetCell{\TADPDA}{1}{deltaJ_rel_median}{\yA}\GetCell{\TADPDB}{1}{deltaJ_rel_median}{\yB}\GetCell{\TADPDC}{1}{deltaJ_rel_median}{\yC}
\addplot[color=cADPD, line width=1.6pt, mark=*]
coordinates {(0.5,\yA) (1.5,\yB) (3.0,\yC)};

\GetCell{\TTripleA}{1}{deltaJ_rel_median}{\yA}\GetCell{\TTripleB}{1}{deltaJ_rel_median}{\yB}\GetCell{\TTripleC}{1}{deltaJ_rel_median}{\yC}
\addplot[color=cTriple, line width=1.6pt, mark=*]
coordinates {(0.5,\yA) (1.5,\yB) (3.0,\yC)};

\nextgroupplot[
  title={Inference noise $\varepsilon=1.50$},
  ytick=\empty,
  yticklabels=\empty,
  extra y ticks={0.0,0.05},
  extra y tick labels={0.00,0.05},
  extra tick style={grid=major}
]

\GetCell{\TNone}{2}{deltaJ_rel_median}{\b}
\addplot[black, dashed, line width=2.5pt, mark=none]
coordinates {(0.5,\b) (1.5,\b) (3.0,\b)};

\GetCell{\TDepA}{2}{deltaJ_rel_median}{\yA}\GetCell{\TDepB}{2}{deltaJ_rel_median}{\yB}\GetCell{\TDepC}{2}{deltaJ_rel_median}{\yC}
\addplot[color=cDep, line width=1.6pt, mark=*]
coordinates {(0.5,\yA) (1.5,\yB) (3.0,\yC)};

\GetCell{\TADA}{2}{deltaJ_rel_median}{\yA}\GetCell{\TADB}{2}{deltaJ_rel_median}{\yB}\GetCell{\TADC}{2}{deltaJ_rel_median}{\yC}
\addplot[color=cAD, line width=1.6pt, mark=*]
coordinates {(0.5,\yA) (1.5,\yB) (3.0,\yC)};

\GetCell{\TPDA}{2}{deltaJ_rel_median}{\yA}\GetCell{\TPDB}{2}{deltaJ_rel_median}{\yB}\GetCell{\TPDC}{2}{deltaJ_rel_median}{\yC}
\addplot[color=cPD, line width=1.6pt, mark=*]
coordinates {(0.5,\yA) (1.5,\yB) (3.0,\yC)};

\GetCell{\TDepADA}{2}{deltaJ_rel_median}{\yA}\GetCell{\TDepADB}{2}{deltaJ_rel_median}{\yB}\GetCell{\TDepADC}{2}{deltaJ_rel_median}{\yC}
\addplot[color=cDepAD, line width=1.6pt, mark=*]
coordinates {(0.5,\yA) (1.5,\yB) (3.0,\yC)};

\GetCell{\TDepPDA}{2}{deltaJ_rel_median}{\yA}\GetCell{\TDepPDB}{2}{deltaJ_rel_median}{\yB}\GetCell{\TDepPDC}{2}{deltaJ_rel_median}{\yC}
\addplot[color=cDepPD, line width=1.6pt, mark=*]
coordinates {(0.5,\yA) (1.5,\yB) (3.0,\yC)};

\GetCell{\TADPDA}{2}{deltaJ_rel_median}{\yA}\GetCell{\TADPDB}{2}{deltaJ_rel_median}{\yB}\GetCell{\TADPDC}{2}{deltaJ_rel_median}{\yC}
\addplot[color=cADPD, line width=1.6pt, mark=*]
coordinates {(0.5,\yA) (1.5,\yB) (3.0,\yC)};

\GetCell{\TTripleA}{2}{deltaJ_rel_median}{\yA}\GetCell{\TTripleB}{2}{deltaJ_rel_median}{\yB}\GetCell{\TTripleC}{2}{deltaJ_rel_median}{\yC}
\addplot[color=cTriple, line width=1.6pt, mark=*]
coordinates {(0.5,\yA) (1.5,\yB) (3.0,\yC)};

\end{groupplot}

\node at ($(dj c1r3.south) + (0,-0.95cm)$) {\huge Training noise level};

\node at ($(dj c1r3.south) + (0,-2.5cm)$) {\pgfplotslegendfromname{RelLegendDeltaJ}};

\node[rotate=90, anchor=south]
at ($(dj c1r1.west)!0.5!(dj c1r3.west) + (-1.35cm,0)$)
{\huge Median relative variation $\Delta \mathsf{J}$};

\end{tikzpicture}

%% file: pgfplots/fig_principal_angle_heatmap.tex

\begin{tikzpicture}

\begin{groupplot}[
  group style={group size=3 by 1, horizontal sep=2mm, group name=hm},
  width=0.31\textwidth,
  height=9.5cm,
  colormap/viridis,
  point meta min=0.002868,
  point meta max=0.143456,
  xmin=-0.5, xmax=3.5,
  xtick={0,1,2,3},
  xticklabels={0.00,0.50,1.50,3.00},
  xlabel={Training noise level},
  ymin=-0.5, ymax=7.5,
  ytick={0,...,7},
  yticklabels={None,Dep,AD,PD,Dep+AD,Dep+PD,AD+PD,Triple},
  y dir=reverse,
  xticklabel style={font=\normalsize},
  yticklabel style={font=\Large},
  label style={font=\Large},
  title style={font=\Large},
  minor tick num=1,
  grid=both,
  major grid style={draw=white, line width=1.2pt},
  minor grid style={draw=white, line width=1.2pt},
]

\nextgroupplot[title={$\varepsilon_0$ = 0.00}]
\DrawInvalidCells
\HeatPanelRow{0}

\nextgroupplot[title={$\varepsilon_0$ = 0.50}, yticklabels={}]
\DrawInvalidCells
\HeatPanelRow{1}

\nextgroupplot[title={$\varepsilon_0$ = 1.50}, yticklabels={}]
\DrawInvalidCells
\HeatPanelRow{2}

\end{groupplot}

\begin{axis}[
  hide axis,
  scale only axis,
  width=0pt,
  height=0pt,
  colormap/viridis,
  point meta min=0.002868,
  point meta max=0.143456,
  colorbar horizontal,
  colorbar style={
      at={(hm c2r1.south)},
      anchor=north,
      yshift=-1.35cm,
      width=0.85\textwidth,
      height=0.35cm,
      xlabel={\Large Mean principal angle (radians)},
      xtick={0.02,0.04,0.06,0.08,0.10,0.12,0.14},
      minor tick num=0,
      tick label style={
        /pgf/number format/fixed,
        /pgf/number format/precision=2,
        font=\small
     }
    }
]
\addplot[draw=none] coordinates {(0,0) (1,1)};
\end{axis}

\end{tikzpicture}

%% file: pgfplots/fig_alignment.tex
\begin{tikzpicture}

\def\aligndir{pgfplots/csv_jacobians}

\begin{axis}[
    width=12.5cm,
    height=7.5cm,
    xlabel={Training noise strength},
    ylabel={Alignment Index},
    grid=both,
    grid style={opacity=0.25},
    tick label style={font=\large},
    label style={font=\Large},
    legend style={
        at={(0.5,-0.22)},
        anchor=north,
        legend columns=4,
        draw,
        font=\normalsize,
        },
    xtick={0.0,0.5,1.0,1.5,2.0,2.5,3.0},
    xticklabels={0.0,0.5,1.0,1.5,2.0,2.5,3.0},
    ytick={0.01,0.02,0.03,0.04,0.05,0.06,0.07},
    yticklabels={0.01,0.02,0.03,0.04,0.05,0.06,0.07},
    tick label style={font=\large},
    scaled y ticks=false,
]

\newcommand{\PlotAlignSeries}[3]{%
  \addplot[
    line width=1.4pt,
    color=#3,
    mark=none
  ]
  table[x index=0, y index=1] {\aligndir/#1_curve.dat};
  \addlegendentry{#2}

  \addplot[
    only marks,
    mark=*,
    mark size=2.2pt,
    draw=#3,
    fill=#3,
    forget plot
  ]
  table[x index=0, y index=1] {\aligndir/#1_curve.dat};

  \addplot[
    line width=1.4pt,
    dashed,
    opacity=0.55,
    color=#3,
    mark=none,
    forget plot
  ]
  table[x index=0, y index=1] {\aligndir/#1_baseline.dat};
}

\PlotAlignSeries{AD}{AD}{cAD}
\PlotAlignSeries{AD+PD}{AD+PD}{cADPD}
\PlotAlignSeries{Dep}{Dep}{cDep}
\PlotAlignSeries{Dep+AD}{Dep+AD}{cDepAD}
\PlotAlignSeries{Dep+PD}{Dep+PD}{cDepPD}
\PlotAlignSeries{PD}{PD}{cPD}
\PlotAlignSeries{Triple}{Triple}{cTriple}

\end{axis}
\end{tikzpicture}